\documentclass[12pt,titlepage,a4paper]{article}
\usepackage{fancyheadings}
\usepackage{epsfig}
\usepackage{amssymb}
\usepackage{amsmath}
\usepackage{amsfonts}
\usepackage{color,graphicx}

\title {\sc {\bf Joint Entropy Coding and Encryption \\using Robust Chaos}}
\author{Nithin Nagaraj, Prabhakar G Vaidya, Kishor G Bhat \\ \\ School of Natural and Engineering Sciences \\ National Institute of Advanced Studies\\ {\bf
nithin\_nagaraj@yahoo.com}}
\date{August 14, 2006}
\begin{document}
\maketitle
\begin{abstract}
We propose a framework for joint entropy coding and encryption
using Chaotic maps. We begin by observing that the message symbols
can be treated as the symbolic sequence of a discrete dynamical
system. For an appropriate choice of the dynamical system, we
could back-iterate and encode the message as the initial condition
of the dynamical system. We show that such an encoding achieves
Shannon's entropy and hence optimal for compression. It turns out
that the appropriate discrete dynamical system to achieve
optimality is the piecewise-linear Generalized Lur\"{o}th Series
(GLS) and further that such an entropy coding technique is exactly
equivalent to the popular Arithmetic Coding algorithm. GLS is a
generalization of Arithmetic Coding with different modes of
operation.

\par GLS preserves the Lebesgue measure and is ergodic. We show that these
properties of GLS enable a framework for joint compression and
encryption and thus give a justification of the recent work of
Grangetto {\it et al.} and Wen {\it et al.} Both these methods
have the obvious disadvantage of the key length being equal to the
message length for strong security. We derive measure preserving
piece-wise non-linear GLS (nGLS) and their skewed cousins, which
exhibit {\it Robust Chaos}. We propose a joint entropy coding and
encryption framework using skewed-nGLS and demonstrate Shannon's
desired sensitivity to the key parameter. Potentially, our method
could improve the security and key efficiency over Grangetto's
method while still maintaining the total compression ratio. This
is a new area of research with promising applications in
communications.
\end{abstract}

\section{Introduction}
The source coding problem is simple to state: given a source $X$
which is emitting bits of information in the absence of noise,
what is the shortest possible way to represent this information?
Stated equivalently, how do we achieve the best possible {\it
compression} of data emitted by a source.

\par Shannon~\cite{Shannon48} gave the limit of ultimate data
compression by introducing the concept of {\it entropy}. Shannon's
entropy of a source is defined as the amount of information
content or the amount of uncertainty associated with the source or
equivalently, the least number of bits per symbol required to
represent the information content of the source without any loss.
Shannon did provide a method (Shannon-Fano coding~\cite{Salomon})
which achieves this limit as the block-length (number of symbols
taken together) for coding increases asymptotically to infinity.
Huffman~\cite{Huffman} provided what are called minimum-redundancy
codes with integer code-word lengths and which achieve Shannon's
Entropy in the limit of the block-length tending to infinity.
However, there are problems associated with both Shannon-Fano
coding and Huffman coding. As the block-length increases, the
number of alphabets exponentially increases, thereby increasing
the memory needed for storing and handling. Also, the complexity
of the encoding algorithm increases since these methods build
code-words for all possible messages of a given length.

\par In this paper, we address the source coding problem from a
different perspective. Since most sources in nature are {\it
non-linear}, we model the information bits of the source $X$ as
measurement bits of a non-linear dynamical system. We treat the
bits of information as the symbolic sequence of a non-linear
dynamical system. For purposes of simplicity and universality, we
want our non-linear dynamical system to be discrete and piece-wise
linear. The simplest such system is the Tent map~\cite{Aligood}.

\par In the next section, we show how we can use the Tent map to
encode binary messages. However, we do not achieve optimality with
the standard Tent map. We discuss a method to achieve optimality.
We show that this leads us to the Generalized Lur\"{o}th Series
(GLS) map and surprisingly turns out that we have re-discovered
the popular Arithmetic Coding algorithm. In Section 4, we discuss
the problem of joint compression and encryption. We briefly review
recent work in this direction in Section 5 and point out their
drawbacks. We aim to provide a motivation for why GLS is a good
framework for joint compression and encryption in Section 6. We
then derive the equations for measure-preserving piecewise
non-linear GLS (nGLS) in Section 7 and discuss the most important
feature of these maps namely ``Robust Chaos'' in Section 8. We
then derive their skewed cousins (skewed-nGLS) in Section 9. In
Section 10, we indicate how skewed-nGLS may be used in joint
entropy coding and encryption and demonstrate Shannon's desired
sensitivity to the key. We also discuss potential advantages of
our method and implications on compression efficiency in the same
section. We summarize our work with future research directions in
Section 11.
\section{Entropy coding using the Chaotic Tent Map}
\par
\begin{figure}[!hbp]
\centering
\includegraphics[scale=.6]{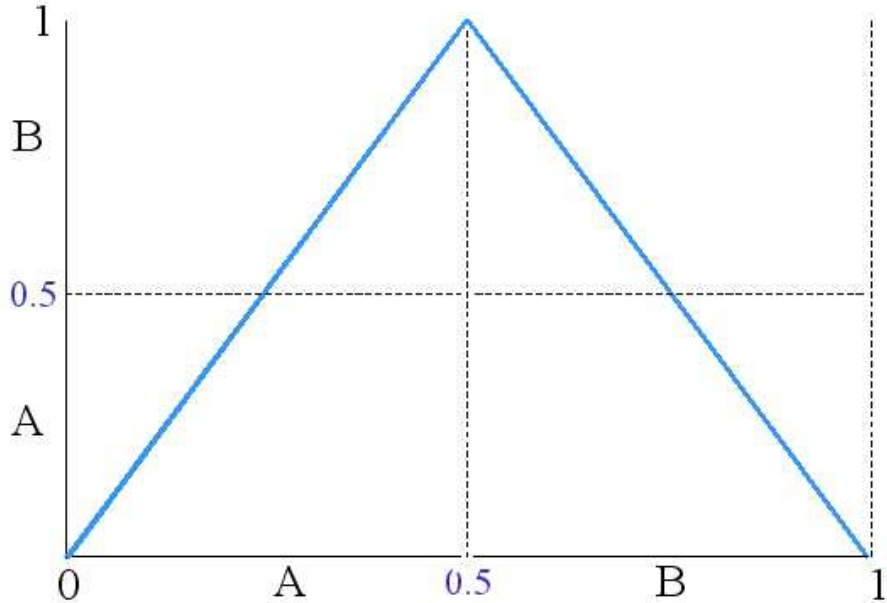}
\caption{The Tent Map.} \label{fig:figtent1}
\end{figure}

 We shall now demonstrate how we can use the Tent map,
one of the simplest chaotic maps to encode a message. Consider the
message $M=`AABABBABAA'$ of length $N=10$ bits. We have two
partitions in the Tent map, the one pertaining to the alphabet $A$
is $[0,0.5]$ and the other interval $(0.5,1]$ corresponds to $B$.
We have the same partitions on the y-axis, as shown in
Figure~\ref{fig:figtent1}. The map consists of linear mappings
from the two partitions to $[0,1]$:
\begin{eqnarray*}
y & = & 2x~~~~~~~~ 0 \leq x < 0.5 \\
&=& 2-2x~~~ 0.5 \leq x \leq 1 \label{eqn:tent}
\end{eqnarray*}
\begin{figure}[!hbp]
\centering
\includegraphics[scale=.6]{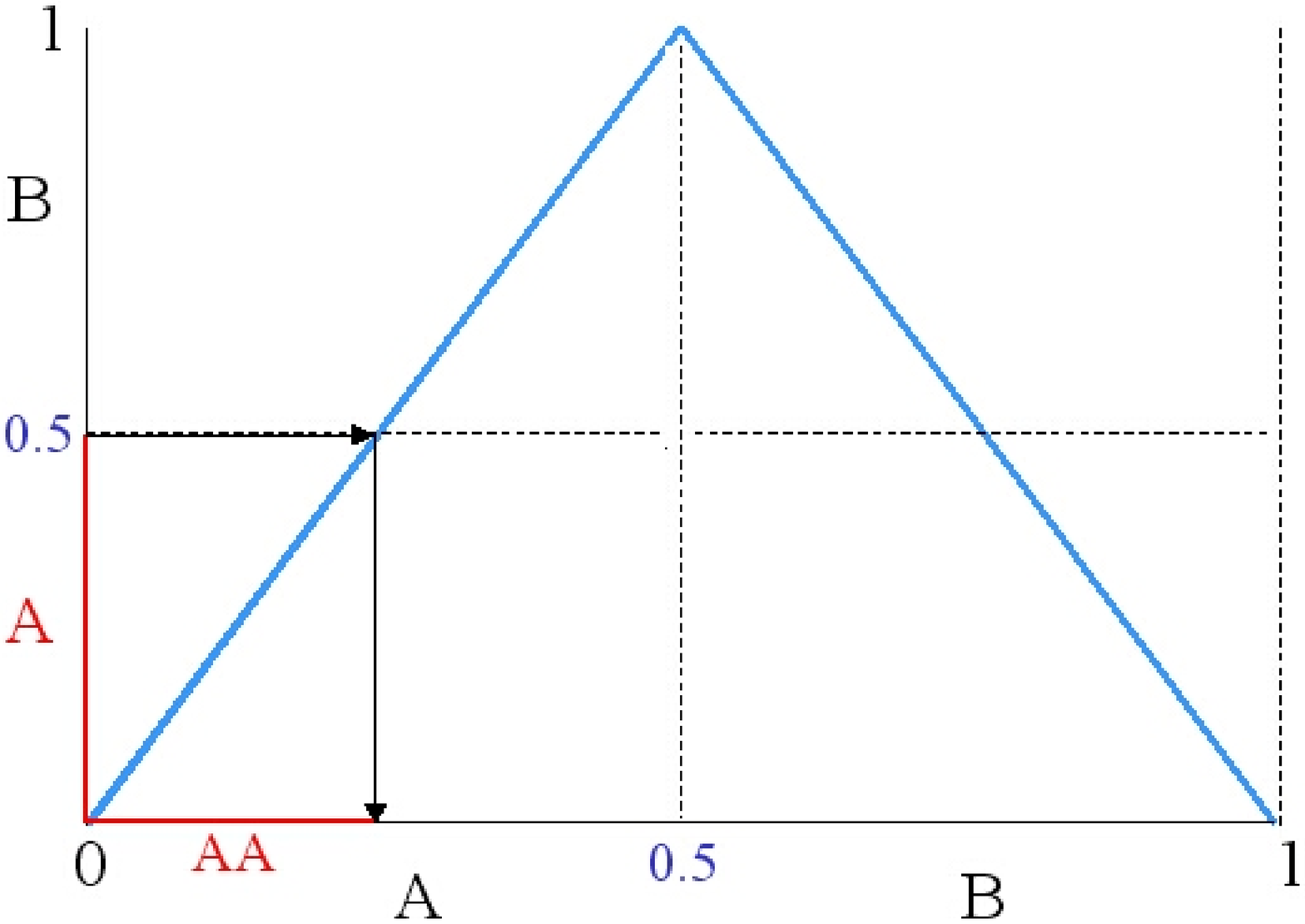}
\caption{Back iteration on the Tent map to encode message
$M=`AABABBABAA'$.} \label{fig:figtent2}
\end{figure}
\begin{figure}[!hbp]
\centering
\includegraphics[scale=.6]{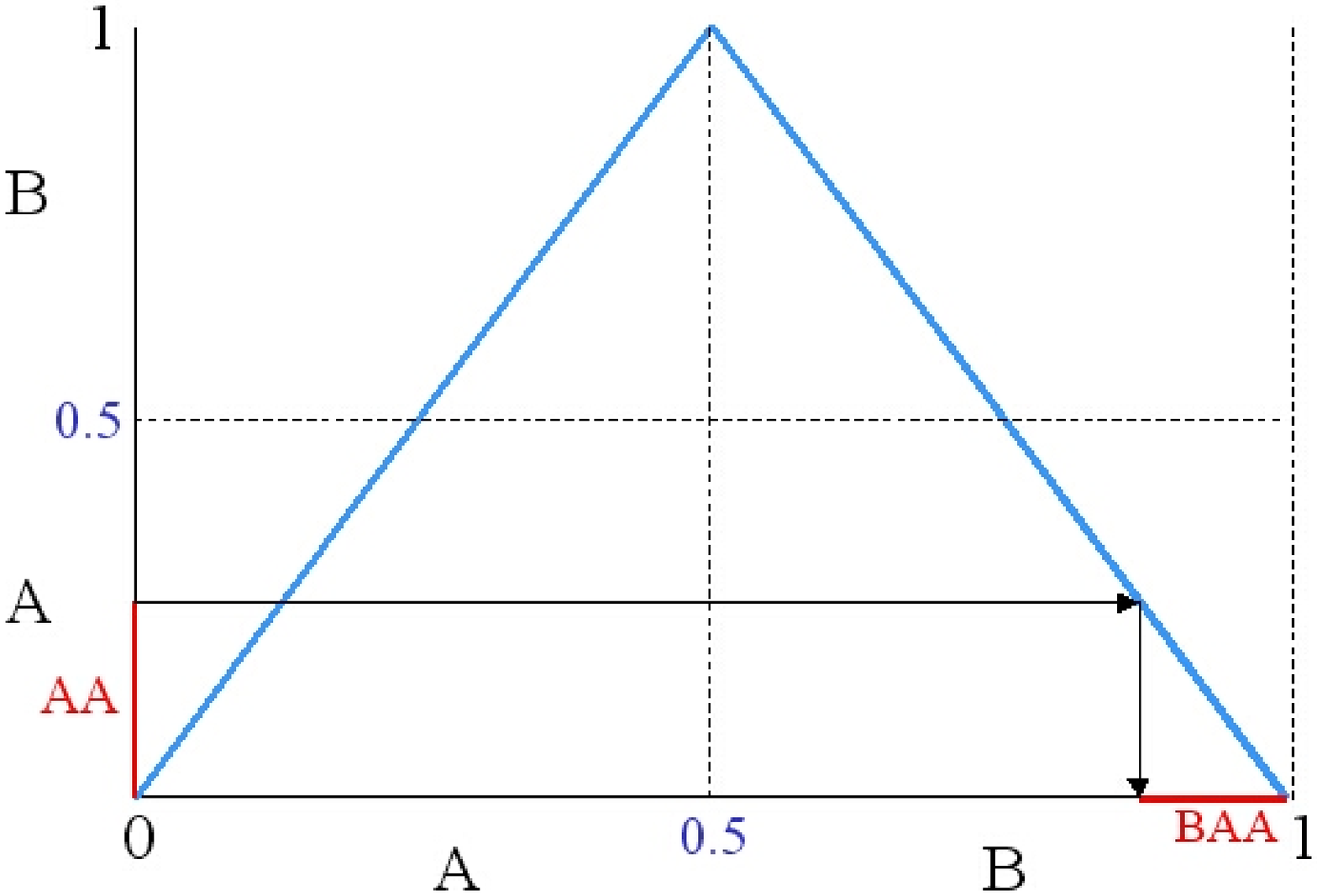}
\caption{Back iteration on the Tent map to encode message
$M=`AABABBABAA'$.} \label{fig:figtent3}
\end{figure}
\begin{figure}[!hbp]
\centering
\includegraphics[scale=.6]{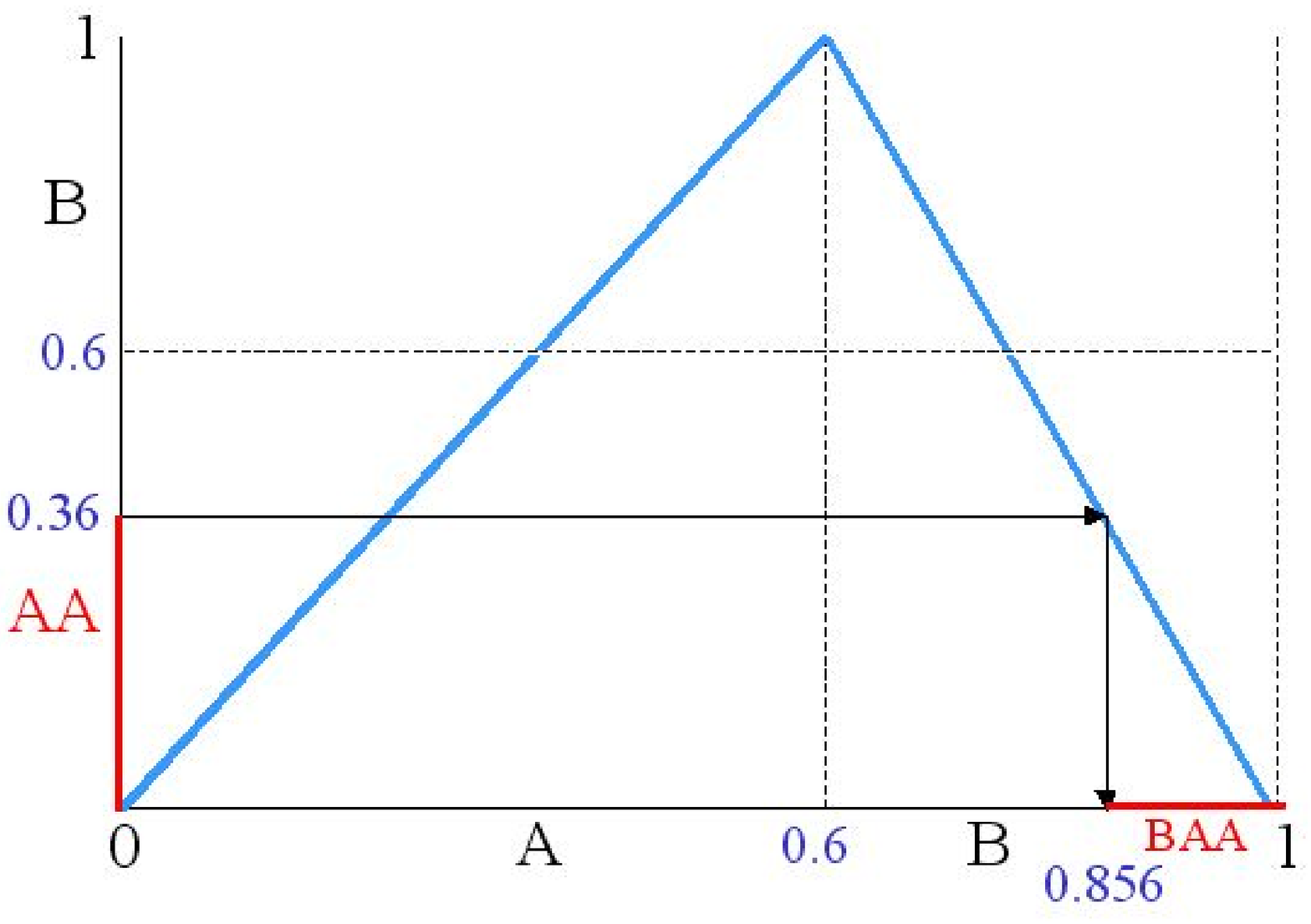}
\caption{Back iteration on the skewed Tent map to encode message
$M=`AABABBABAA'$.} \label{fig:figsktent1}
\end{figure}
\par To begin coding the message $M=`AABABBABAA'$, we begin from
the last symbol of the message and back iterate
(Figure~\ref{fig:figtent2}). Since the last symbol is $A$, we
begin with the partition $[0,0.5)$ on the y-axis and look at its
pre-image. There are two pre-images of this interval corresponding
to the two linear maps. Since the previous symbol is also $A$, we
take the first pre-image. This would correspond to $[0,0.25)$. We
then compute the next back iterate. The symbol is $B$ and hence we
take the second pre-image of $[0,0.25)$ which is (0.875, 1] as
shown in Figure~\ref{fig:figtent3}. We keep back-iterating in this
fashion until we finally stop (since the message has finite
length, this process has to terminate). We end up with an interval
$[START, END]$, inside which our initial condition is going to
lie. We could choose any real number in this interval as the
initial condition. For the sake of simplicity, we choose the
mid-point $\frac{START+END}{2}$ as the initial condition. This
initial condition is binary coded and transmitted to the decoder.

\par What we have done effectively is that we have treated the message symbols
as the symbolic sequence of the Tent map. We have obtained the
initial condition which the decoder forward iterates to yield a
trajectory, the symbolic sequence of which is the desired message.
Two questions arise:
\begin{enumerate}
\item How many bits of the initial condition needs to be
transmitted to the decoder? %
\item Is such a method optimal in terms of compression? %
\end{enumerate}
\par The answer to the first question is straightforward. It is easy to see that the number of bits
needed to transmit is $\lceil -log_2(END - START) \rceil + 1$ to
uniquely distinguish the message from all other messages of the
same length. This implies that the number of bits we transmit
depends on the length of the final interval ($END-START$). Longer
the message, shorter is the interval and hence more bits need to
be sent. The second question is more important and the answer is
negative. The method is not optimal, in fact, no compression is
achieved by this method. This can be easily seen by observing that
all possible binary messages of length $N$ bits would get an
interval of size $2^{-N}$ units on the real line $[0,1]$ by the
encoding we have just described. The number bits needed to
transmit the initial condition of a $2^{-N}$ long interval is $N$
bits. Hence no compression is achieved.
\subsection{Encoding using the skewed Tent map}
We shall now modify our method to make it Shannon optimal. By
Shannon optimal, we mean the compression  achieved should approach
the Shannon's entropy of the message as the length of the message
increases. We notice that the problem we were having is that the
standard Tent map is treating all messages as if they were equally
likely. This is where the probability model of the source comes
into picture. The Tent map treated both $A$ and $B$ as equally
likely ($p(A)=p(B)=0.5$ where $p(.)$ denotes first-order
probability). This is true only for a perfect random source and we
know that a true random sequence is uncompressible. Since most
real-world messages that we are interested in storage and
transmission are far from random, there is scope for compression.
We modify the Tent map to account for the skew in the
probabilities of A and B. Specifically, we allocate the intervals,
the length of which are {\it equal} to the probability of the
corresponding alphabet. Thus for the particular example
$M=`AABABBABAA'$, we first compute the probabilities of $A$ and
$B$ as shown in Table~\ref{tab:tabprob}. Then allocate the range
$[0,0.6]$ to $A$ and $(0.6,1]$ to $B$. The encoding proceeds
exactly in the same fashion as before (refer to
Figure~\ref{fig:figsktent1}). The decoding is also unchanged.
However, the probability model of the source has to be now
available at the decoder and hence needs to be sent along with the
coded message.
\begin{table}
\centering
\begin{tabular}{|c|c|c|}
  \hline
  Character & Probability & Range \\
  \hline
  A & $\frac{6}{10} = 0.6$ & $[0, 0.6]$ \\
    & &  \\
  B & $\frac{4}{10} = 0.4$ & $(0.6, 1.00]$ \\
  \hline
\end{tabular}
\vspace{0.1in} \caption{Probability model for the example. }
\label{tab:tabprob}
\end{table}
\subsection{Shannon Optimality of the modified method}
\par We shall now address the issue of Shannon's optimality for compression.
For the particular example, our method yields an initial condition
which requires $11$ bits. The Shannon's entropy (first-order) for
the message is computed as $H = -\sum_{i=0}^{i=1}
p_i.log_{2}(p_i)$ where $i=0$ corresponds to the symbol $A$ and
$i=1$ to $B$ and $p_i$ refers to the probability of the i$^{th}$
source alphabet. This is found to be $H=0.971$ bits/symbol. This
means that for a message of length 10 bits, the optimal number of
bits $=10 \times H$ which is $9.71$ bits. We don't seem to achieve
optimality for this example. However, with the same probability
model for a message of length 1000 bits, our method would transmit
972 bits as opposed to the optimal value of 971 bits. Thus one can
see that as the message gets longer, our method approaches
Shannon's optimality.

\par We make the important observation that the Tent map is a type
of Generalized Lur\"{o}th Series (GLS)~\cite{Dajani}. Hence, we
shall call this method GLS entropy coding or GLS coding method. We
shall now prove theoretically that we achieve Shannon's optimal
limit by showing that GLS coding is equivalent to Arithmetic
coding, a popular coding method which is Shannon optimal.
\section{GLS coding is equivalent to Arithmetic Coding and hence Shannon Optimal}
Let us briefly visit Arithemtic Coding (AC). AC is a popular
entropy coding method which has its origins in the early 1960s
(Elias and others). However, it gained wide acceptance after the
1979 paper by Rissanen and Langdon~\cite{Rissanen} who gave a
practical implementation of the method. Today, AC is one of the
most widely used entropy coding methods owing to its optimality
and also improved speed of decoding.

\par The idea of AC is to first give a unique tag to the entire
sequence~\cite{Sayood}. This is unlike Huffman coding which gives
individual codes to symbols of the message. Since AC codes the
entire sequence rather than coding individual symbols, the length
of the code-words may not be integers. The tag is then binary
coded and transmitted. In AC, there is no need to generate codes
for all sequences at a time and hence very efficient for long
sequences. Huffman coding has the disadvantage of having to
generate code-words for all possible messages of a given length.

\par The Real line $[0,1)$ is used to generate tags. AC is Shannon optimal
without the necessity of blocking. As the length of the message
increases, AC comes closer to Shannon's entropy~\cite{Sayood}.
\subsection{A Binary Example}
We shall illustrate the coding method of AC on the same message.
\begin{figure}[!hbp]
\centering
\includegraphics[scale=.6]{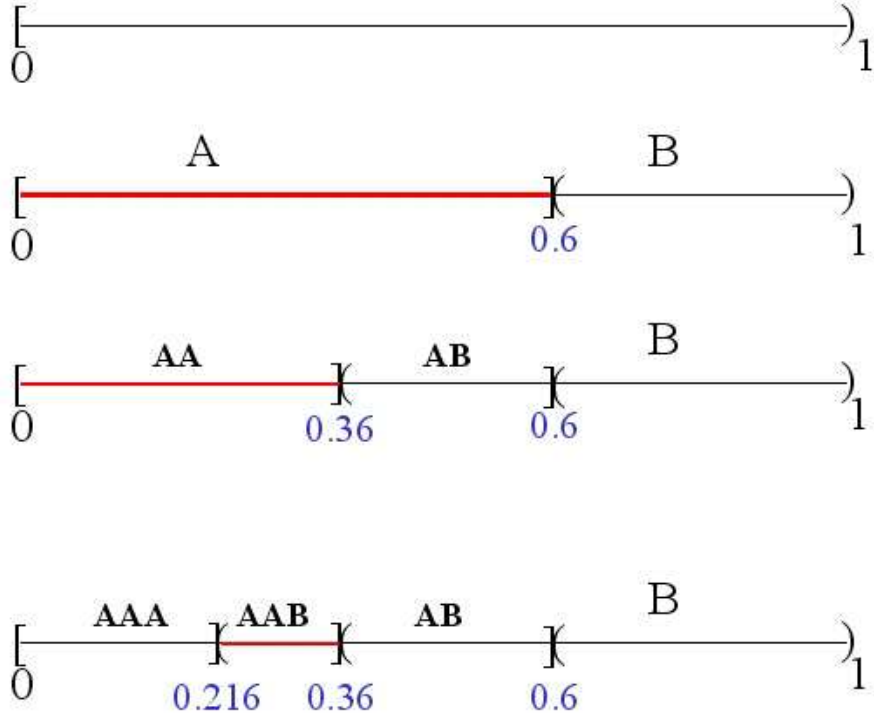}
\caption{Few iterations of AC encoding of the message
$M=`AABABBABAA'$} \label{fig:figaceg}
\end{figure}
\begin{figure}[!hbp]
\centering
\includegraphics[scale=.6]{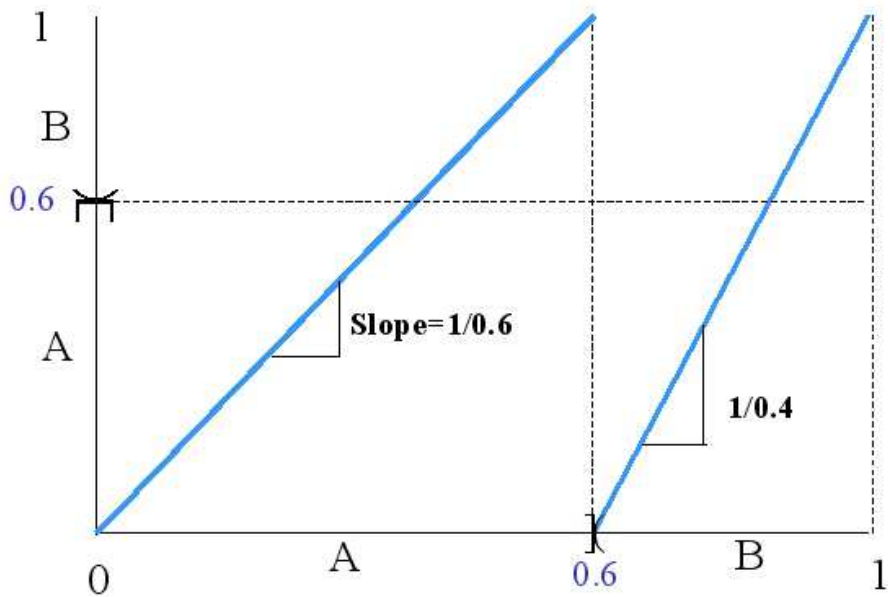}
\caption{The mode of Generalized Lur\"{o}th Series (GLS) which is
exactly the same as AC. Both methods yield identical intervals for
encoding.} \label{fig:figgls1}
\end{figure}
First compute the probabilities of $A$ and $B$ as shown in
Table~\ref{tab:tabprob}. Then allocate the range $[0,0.6]$ to $A$
and $(0.6,1)$ to $B$. In order to encode the message $M$, we
observe that the first symbol is $A$ and hence the tag will lie in
$(0,0.6]$ (refer to the red line marked in
Figure~\ref{fig:figaceg}). We subdivide the interval $[0,0.6]$
into two parts in the ratio $0.6:0.4$, allocating the left one
$[0,0.36]$ to $AA$ and $(0.36,0.6]$ to $AB$. Since the second
symbol is $A$, the tag will lie in the interval allocated to $AA$
which is $[0,0.36]$. The third symbol is $B$ and so we sub-divide
the interval corresponding to $AA$ into two parts in the same
ratio ($0.6:0.4$) allocating the left one to $AAA$ ($[0,0.216)$)
and the right one to $AAB$ ($(0.216,0.36]$). The tag is going to
lie inside $AAB$. We proceed along the same lines until we finally
stop (since the message has finite length, this process has to
terminate). We end up with an interval $[START, END]$, inside
which the tag lies. We could choose any real number in this
interval as a tag. For the sake of simplicity, we choose the
mid-point $\frac{START+END}{2}$ as the tag. This tag is binary
coded and transmitted to the decoder.
\par We claim that the length of the interval obtained in the above described traditional AC
coding is the same as the length of the interval in GLS coding. To
see this, we notice that at every iteration of the GLS, the length
of the interval we started with is multiplied by the probability
of the symbol being encoded to yield us the new length. Thus, at
the end of the iterations, the length of the final interval will
be $p(A)^k p(B)^{N-k}$ where $p(.)$ denotes the first-order
probability of the alphabets, `$k$' is the number of $A$-s in the
message and $N-k$ is the number of $B$-s in the message for a $N$
bit length message. This is exactly the probability of the message
(treating each symbol as independent) and is also the length of
the interval for AC. Since the lengths of the final intervals in
both AC and GLS coding are the same, the number of bits needed to
encode the initial condition will be identical, thus yielding
exactly the same compression ratio. Since AC is Shannon optimal,
GLS is also Shannon optimal.

\par It actually turns out that there are different modes of the GLS (which we shall see later)
and one of them corresponds to AC. This means that there exists a
particular mode of GLS (Figure~\ref{fig:figgls1}) where not only
the length of the final interval is matched with AC, but also the
exact interval itself. Note that all modes of GLS are Shannon
optimal. Hence, GLS coding can be thought of as a generalization
of the AC coding and achieves Shannon's optimality of compression
efficiency.

\par It is important for us to acknowledge Luca's work~\cite{Luca}
in this context. Luca claims a new method of entropy coding using
a chaotic map very similar to ours (their map is exactly the same
as AC). However, Luca doesn't seem to realize that their method is
essentially an alternate narrative of AC. They do not make the
observation that it is a GLS with a different mode of operation.
However, it is important to acknowledge Luca's contribution in
being able to see the coding operation as a back-iteration on an
one-dimensional chaotic map (the act of seeing the second
dimension in the coding operation is a key thing which they do in
their paper).
\section{Joint Coding and Encryption: The Problem Statement}
The problem that we are now interested in is the following: how to
transmit information {\it efficiently} and {\it securely} from
point $A$ to point $B$?
\begin{figure}[!hbp]
\centering
\includegraphics[scale=.6]{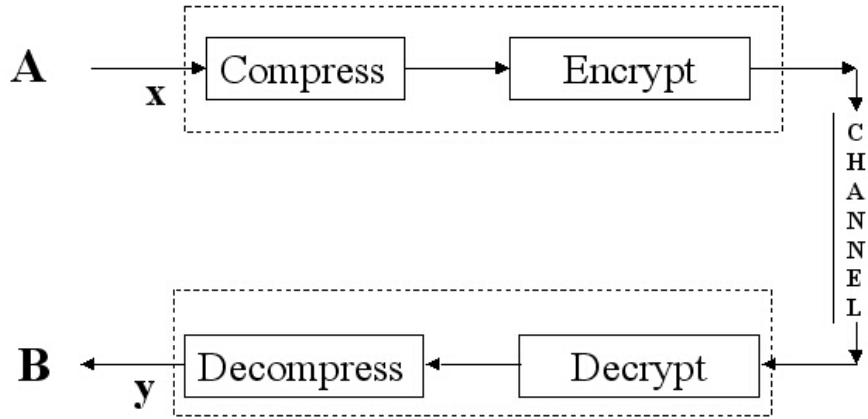}
\caption{The problem statement.} \label{fig:figprob1}
\end{figure}
\par In order to achieve the above objective, we have to subject
the message $x$ to compression in order to come up with a
parsimonious representation of its information content. We wish to
transmit as little as possible and we know that the best lossless
compression theoretically achievable is bounded by Shannon's
Entropy from below. Since we also wish to transmit message $x$
securely, we need to disguise this information by encryption. At
the receiver end, we need the corresponding blocks of decryption
and decompression to recover the message $y$ (refer to
Figure~\ref{fig:figprob1}). We make the following assumptions:
\begin{enumerate}
    \item Noiseless source and channel:~There is no noise either at the
    source or at the channel. We will not be needing channel
    codes.
    \item Lossless coding:~ All coding will be lossless which
    implies $x=y$. We shall use the word `coding' to imply
    compression and `decoding' to imply decompression throughout this paper.
    \item Eavesdroppers:~ There are eavesdroppers on the channel.
\end{enumerate}
\begin{figure}[!h]
\centering
\includegraphics[scale=.6]{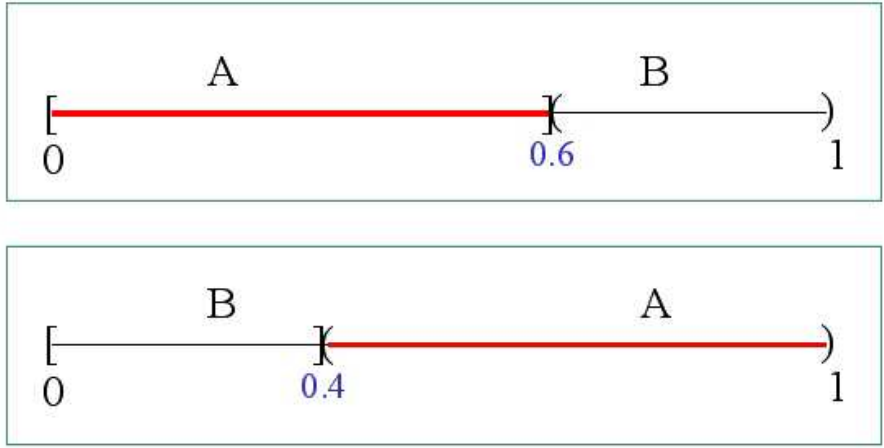}
\caption{Grangetto's method~\cite{Grangetto}: swap between the two
modes based on a random key.} \label{fig:figgrang}
\end{figure}
\section{Joint Entropy Coding and Encryption: Previous Work}
In this section, we shall briefly review some previous work on
joint entropy coding and encryption.

\par To the best of our knowledge, there are only two frameworks for joint entropy coding and
encryption, that of Grangetto~\cite{Grangetto} and Wen~\cite{Wen},
and both use AC as the entropy coding algorithm. Grangetto
proposed {\it randomized AC} where at every coding iteration, the
intervals of the binary alphabet were switched (or not-switched)
based on a random key (Figure~\ref{fig:figgrang}).

\par This random switching of the two intervals has the effect of
randomizing the {\it location} of the final interval $[START,
END]$ in which the tag is going to lie. The key consists of 1 bit
per coding iteration and hence is essentially as long as the
message itself. The key is assumed to be transmitted on a secure
channel before the decoding can begin. It can be seen that there
is no loss of optimality with respect to the compression ratio in
this method.

\par Wen's method is a little more complicated where they used
key-based interval splitting so that now the intervals allocated
to symbols at every iteration are no longer contiguous. The key in
this case is also as long as the message. There is a slight loss
in optimality of compression ratio which is negligible for long
sequences.

\par Some draw-backs of both methods are as follows:
\begin{enumerate}
\item Key-distribution problem:~ The key is as long as the message. %
\item Why should the method work, if it works? %
\item In particular, why should AC be a good choice for such a
joint coding and encryption framework? Why not other entropy coding methods like Huffman coding, Shannon-Fano coding etc.? %
\item The length of the final interval $[START, END]$ in which the
TAG lies is not changed by swapping, only its location on the real
line is randomized. This is an important observation which we
shall allude to later(``a good disguise should hide one's
height.'').
\end{enumerate}
\par We hope to provide some answers to the above question and
also propose a method which has the potential to circumvent some
of the above mentioned draw-backs.
\begin{figure}[!h]
\centering
\includegraphics[scale=.6]{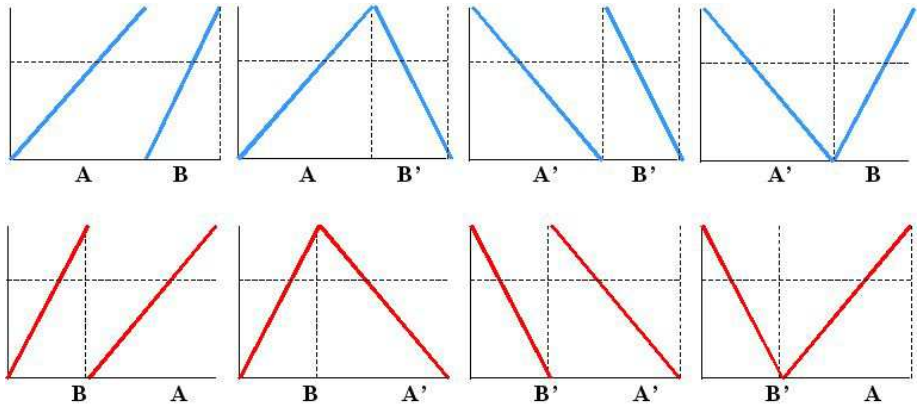}
\caption{Different modes of GLS. All of these are chaotic, ergodic
and Lebesgue measure preserving maps. Grangetto's method involves
swapping between the first mode and the fifth mode (numbering from
left to right, top to bottom).} \label{fig:figgls4}
\end{figure}
\section{Why GLS?}
\par In this section, we wish to provide an answer as to why we
think that GLS is potentially a very good candidate for joint
entropy coding and encryption.
\subsection{GLS a Chaotic, Ergodic, Measure-preserving map}
GLS is an ergodic and Lebesgue measure-preserving discrete
dynamical system~\cite{Dajani}. We intend to make use of these
facts for proposing our method of joint coding and encryption.

\par As previously stated, there are different possible modes of GLS for the binary alphabet case and these are shown
in Figure~\ref{fig:figgls4}. The different modes are obtained by
combining the two operations of reversing the map on the
partitions and by swapping the two partitions. Grangetto's method
involves swapping between modes 1 and 5 at every coding iteration
based on a private key.

\par Our treatment can be easily extended to larger alphabets.
Important properties of the GLS are that it is chaotic, ergodic,
Lebesgue measure preserving and Shannon optimal for compression as
shown in Section 3. We shall show how these play an important role
later.

\subsection{Shannon's remarks from his 1949 masterpiece}
We have not yet fully justified why GLS is the ideal candidate for
a joint coding and encryption framework. We have to visit Shannon
for our argument.

\par We cite here Shannon's statements from his famous 1949 paper
on secrecy of communications sytems~\cite{Shannon49} as quoted by
Kocarev~\cite{Kocarev}: {\it ``Good mixing transformations are
often formed by repeated products of two simple non-commuting
operations. Hopf has shown, for example, that pastry dough can be
mixed by such a sequence of operations. The dough is first rolled
out into a thin slab, then folded over, then rolled, and then
folded again, etc. . . . In a good mixing transformation . . .
functions are complicated, involving all variables in a sensitive
way. A small variation of any one (variable) changes (the outputs)
considerably.''} Here we wish to make several observations.
Shannon is talking about mixing transformation for the purposes of
efficient encryption. We believe that he is hinting towards the
notion of {\it ergodicity} when he refers to mixing. Also, {\it
complicated} could mean non-linear and {\it involving all
variables in a sensitive way} could mean chaotic (sensitive
dependence on initial conditions indicated by positive Lyapunov
exponents). What we are hinting is that Shannon is referring to
Chaos and its use in cryptography, 15 years earlier to the coining
of the term.

\par We have shown that AC is nothing but a GLS which is
piece-wise linear, chaotic, ergodic and Lebesgue measure
preserving discrete dynamical system. We know that it is optimal
(achieves Shannon entropy as the length of the message gets longer
and longer). We have also seen how good encryption methods need to
have properties such as mixing (ergodic), complicated functions
(non-linear) and sensitivity to variables (chaotic, positive
Lyapunov exponents) as per Shannon's words. These are best
provided by a discrete dynamical system which is chaotic and
ergodic. Since our goal is to transmit information efficiently and
securely and since both these functions can be achieved by a
discrete dynamical system under chaos, why not use a {\it single}
dynamical system to achieve both? We believe that it is this
philosophy that provides a justification for using GLS (or AC) as
a framework for joint coding and encryption.
\section{nGLS: Measure-preserving Piecewise Non-linear GLS}
\begin{figure}[!hbp]
\centering
\includegraphics[scale=.6]{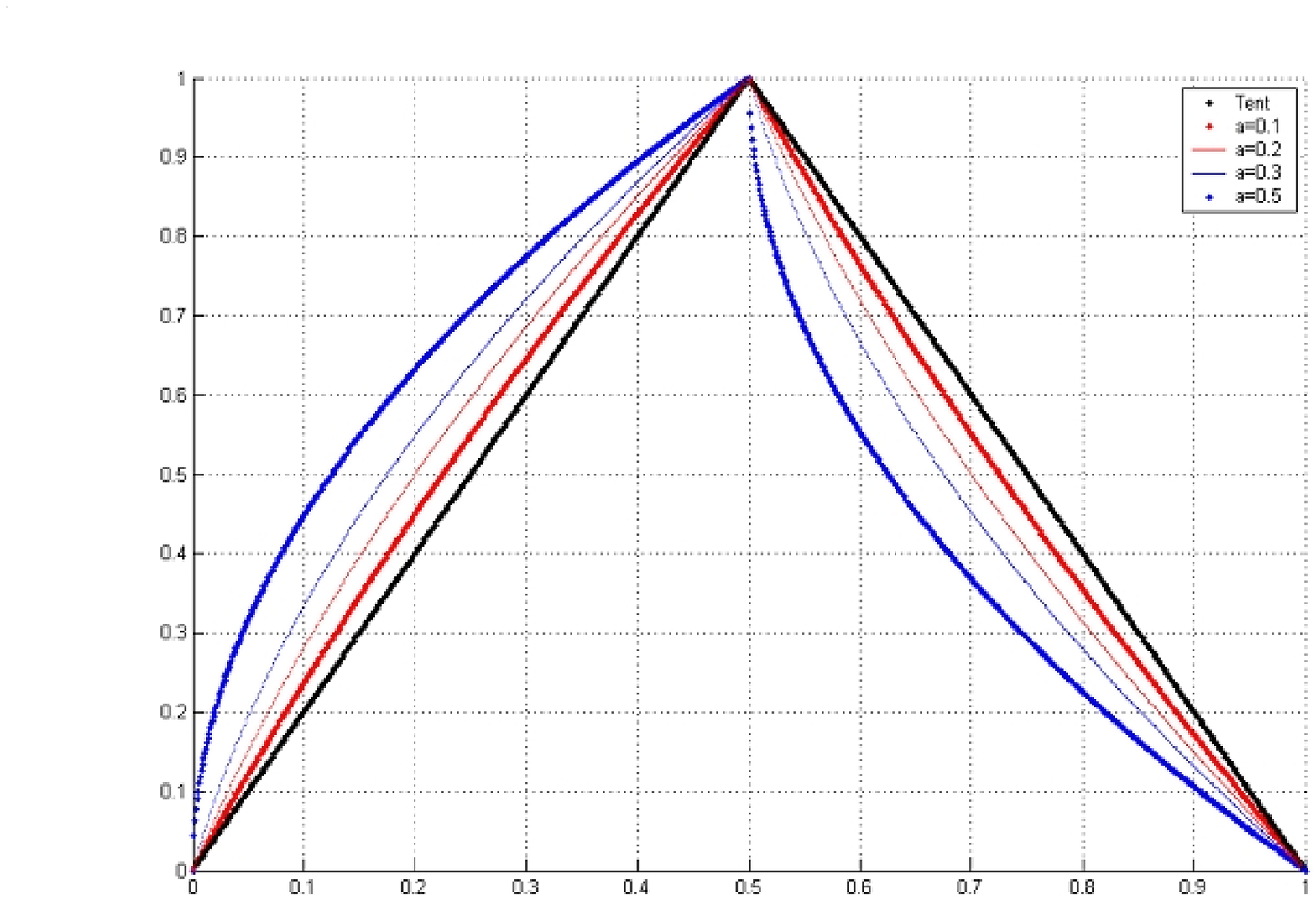}
\caption{nGLS: piece-wise non-linear GLS for various values of
`$a$'.} \label{fig:figngls1}
\end{figure}

In this section, we derive a generalization of GLS which is
piece-wise non-linear. We call this nGLS. To begin with, we shall
consider a binary alphabet standard Tent map.
\begin{eqnarray*}
y &=& 2x  ~~~~~~~~~~ 0 \leq x <0.5\\
  &=& 2-2x ~~~~~ 0.5 \leq x \leq 1
\end{eqnarray*}
We re-write this as:
\begin{eqnarray*}
\frac{1}{2}y &=& x  ~~~~~~~~~~ 0 \leq x <0.5\\
 -\frac{1}{2}y + 1 &=& x ~~~~~~~~~~ 0.5 \leq x \leq 1
\end{eqnarray*}
We can generalize this as:
\begin{eqnarray*}
b_1y + c_1 &=& x  ~~~~~~~~~~ 0 \leq x <0.5\\
b_2y + c_2 &=& x ~~~~~~~~~~ 0.5 \leq x \leq 1
\end{eqnarray*}
where $b_1=\frac{1}{2}$, $c_1 = 0$, $b_2= -\frac{1}{2}$, $c_2=1$.
We add a non-linear term in $y$,
\begin{eqnarray*}
a_1y^2 + b_1y + c_1 &=& x  ~~~~~~~~~~ 0 \leq x <0.5\\
a_2y^2 + b_2y + c_2 &=& x ~~~~~~~~~~ 0.5 \leq x \leq 1
\end{eqnarray*}
We set the constraints $y=0$ at $x=0$, $y=1$ at $x=0.5$ and $y=0$
at $x=1$ and simplify the equations to yield:
\begin{eqnarray*}
ay^2 + (\frac{1}{2}-a)y &=& x  ~~~~~~~~~~ 0 \leq x <0.5\\
ay^2 + (-\frac{1}{2}-a)y + 1&=& x ~~~~~~~~~~ 0.5 \leq x \leq 1
\end{eqnarray*}
We can solve for $y$ to get:
\begin{eqnarray*} y &=&
\frac{-1+2a+\sqrt{1-4a+4a^2+16ax}}{4a}~~~~~~~~~~ 0 \leq x
<0.5\\\\
  &=& \frac{1+2a-\sqrt{1-12a+4a^2+16ax} }{4a}~~~~~~~~~~ 0.5 \leq x \leq 1\\
\end{eqnarray*}
We call the above equations as $nGLS(a,x)$. It is important to
note that as $a \rightarrow 0$, the above equations tend to the
standard Tent map. The $nGLS$ family is plotted for a few values
of `$a$' in Figure~\ref{fig:figngls1}.
\begin{figure}[!hbp]
\centering
\includegraphics[scale=.6]{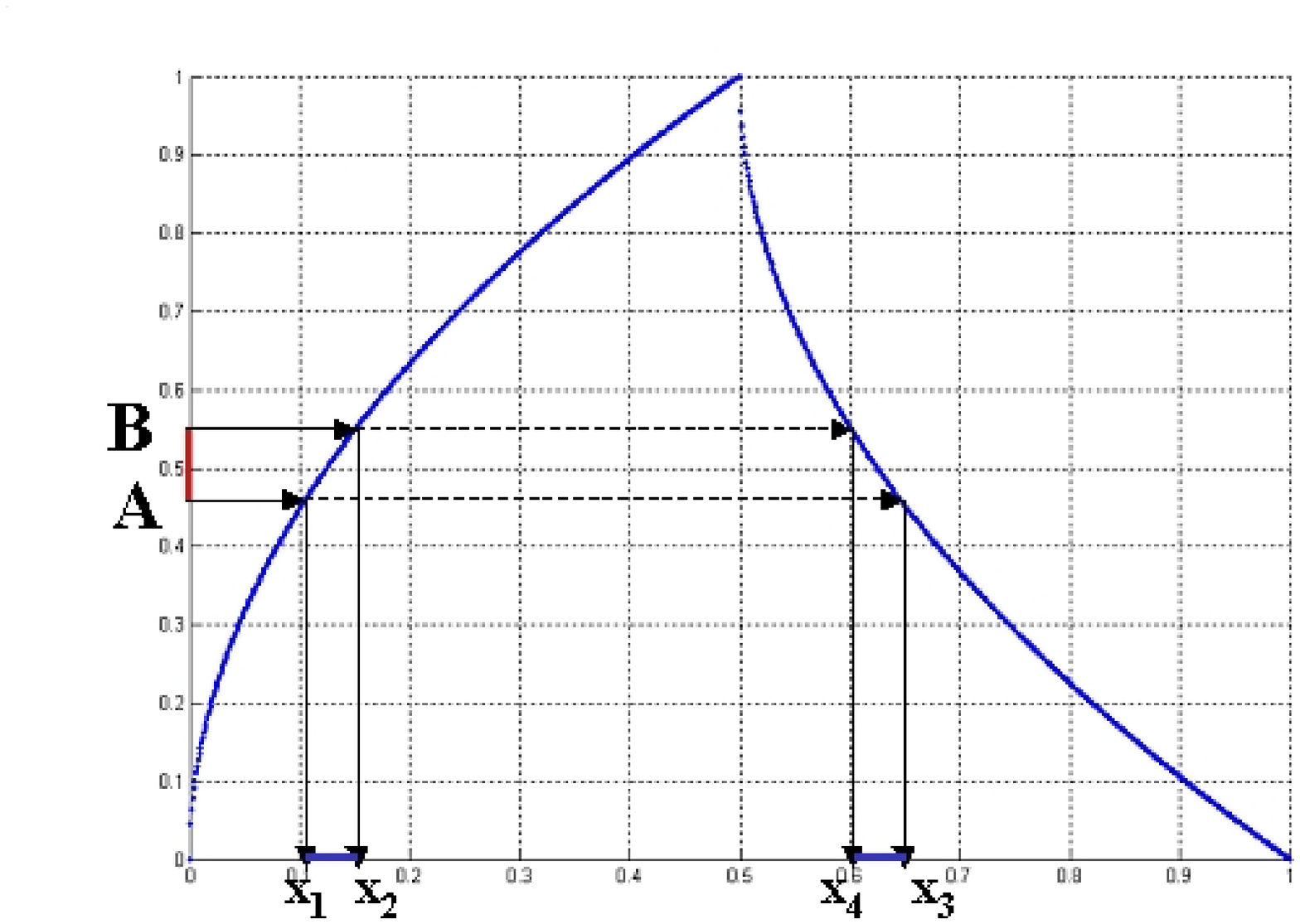}
\caption{nGLS preserves the Lebesgue measure.}
\label{fig:figngls2}
\end{figure}

\subsection{nGLS preserves the Lebesgue measure}
We shall prove that nGLS preserves the Lebesgue measure. In other
words, we need to prove: %
\begin{equation}
 \mu ( nGLS^{-1}([A,B])  ) = \mu ([A, B])  = B-A
\end{equation}
The inverse images of $[A,B]$ are given by
Figure~\ref{fig:figngls2}. These are:
\begin{eqnarray*}
x_1 &=& aA^2 + (\frac{1}{2} - a)A.\\
x_2 &=& aB^2 + (\frac{1}{2} - a)B.\\
x_3 &=& aA^2 + (-\frac{1}{2} - a)A + 1.\\
x_4 &=& aB^2 + (-\frac{1}{2} - a)B + 1.
\end{eqnarray*}
Now,
\begin{eqnarray*}
\mu ( nGLS^{-1}([A,B])  ) &=& (x_2 - x_1) + (x_3 - x_4). \\
&=& a(B^2-A^2) + \frac{B-A}{2} + a(-B+A)\\
& & +a(A^2-B^2) +\frac{B-A}{2} + a(B-A).\\
&=& B-A. \\
&=& \mu ([A, B]).
\end{eqnarray*}
and hence proved.
\subsection{Different modes of nGLS}
Similar to the GLS, there are eight different modes of nGLS which
are all measure-preserving. We omit plotting these modes here.
\begin{figure}[!hbp]
\centering
\includegraphics[scale=.6]{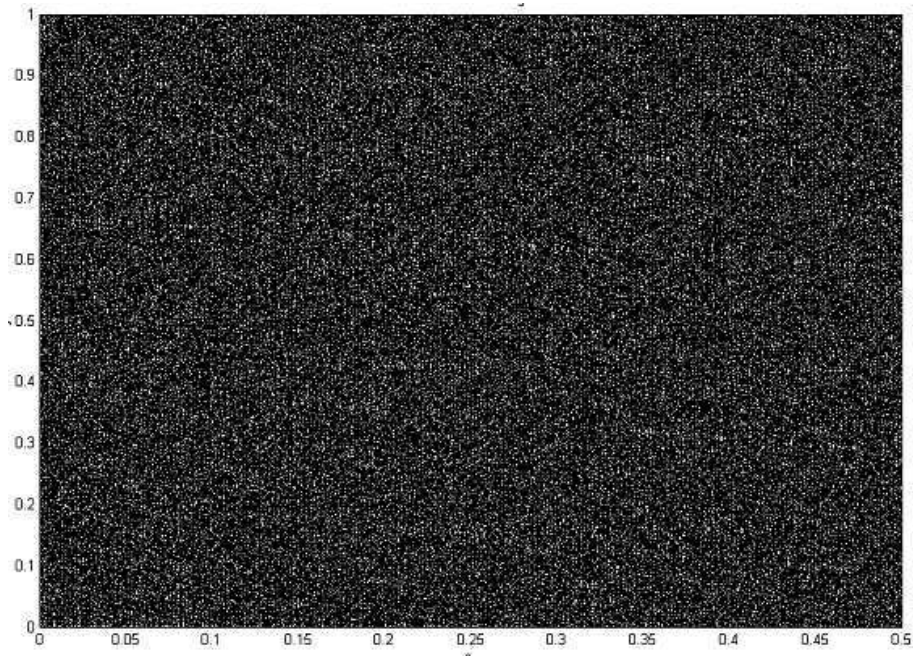}
\caption{nGLS exhibits Robust Chaos in `$a$'. The bifurcation
diagram is shown above for nGLS for $0 < a \leq 0.5$.}
\label{fig:figrobust1}
\end{figure}
\begin{figure}[!hbp]
\centering
\includegraphics[scale=.6]{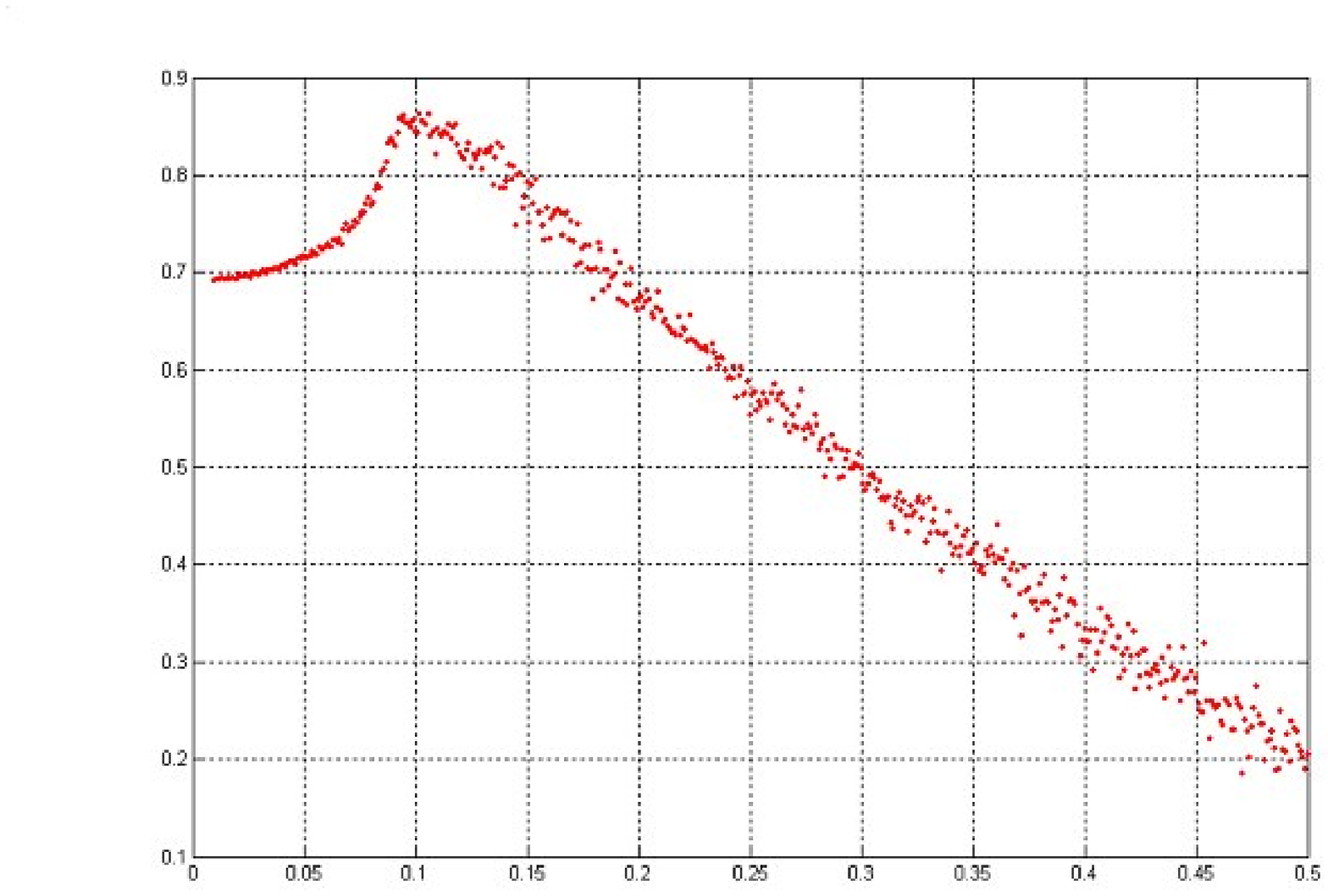}
\caption{Lyapunov exponents of nGLS for $0 < a \leq 0.5$. They are
found to be positive.} \label{fig:figlyap1}
\end{figure}
\section{nGLS exhibits ``Robust Chaos''}
Robust Chaos is defined by the absence of periodic windows and
coexisting attractors in some neighborhood of the parameter
space~\cite{Banerjee}. Barreto~\cite{Barreto} had conjectured that
robust chaos may not be possible in smooth unimodal
one-dimensional maps. This was shown to be false with
counter-examples by Andrecut~\cite{Andrecut} and
Banerjee~\cite{Banerjee}. Banerjee demonstrates the use of robust
chaos in a practical example in electrical engineering. Andrecut
provides a general procedure for generating robust chaos in smooth
unimodal maps.

\par As observed by Andrecut~\cite{Andrecut2}, robust chaos implies
a kind of ergodicity or good mixing properties of the map. This
makes it very beneficial for cryptographic purposes. The absence
of windows would mean that the these maps can be used in hardware
implementation as there would be no {\it fragility} of chaos with
noise induced variation of the parameters. Recently, we have
demonstrated the use of Robust Chaos in generating pseudo-random
numbers which passes rigorous statistical randomness
tests~\cite{Mahesh}.

\par nGLS exhibits Robust Chaos in the parameter `$a$' as inferred
from the bifurcation diagram in Figure~\ref{fig:figrobust1}.
\subsection{Positive Lyapunov exponents}
The Lyapunov exponent is experimentally determined for the
parameter $0 < a \leq 0.5$ and is found to be positive. This is a
necessary condition for chaos. Figure~\ref{fig:figlyap1} shows a
plot of Lyapunov exponents for nGLS for the bifurcation parameter
$0 < a \leq 0.5$. They are found to be positive.
\section{Skewed-nGLS}
\begin{figure}[!hbp]
\centering
\includegraphics[scale=.6]{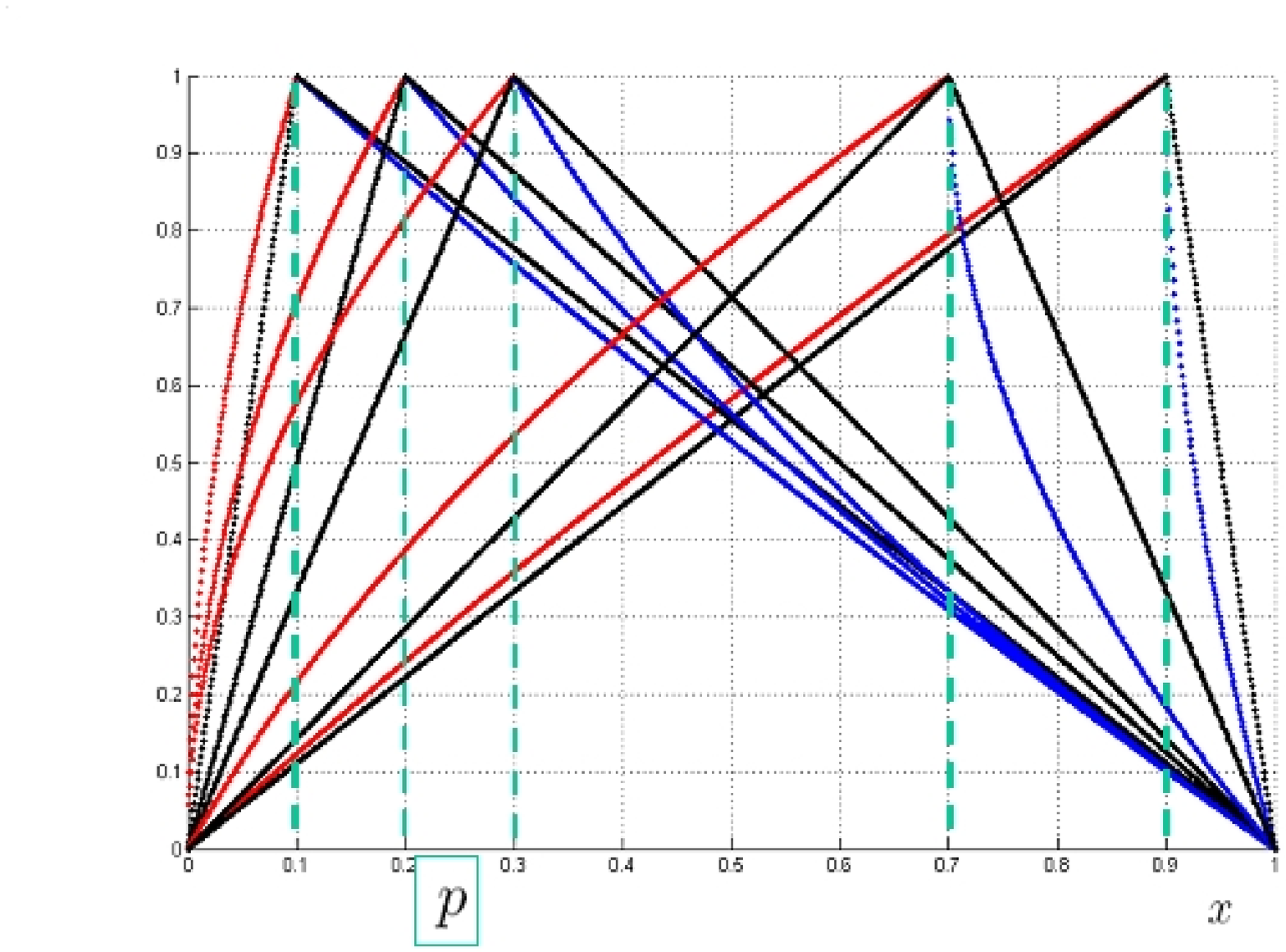}
\caption{Skewed-nGLS vs. x.} \label{fig:figsngls1}
\end{figure}

We have so far only considered the symmetric case (the partitions
$A$ and $B$ are equal). We now, do similar analysis with a skew
and arrive at the following family of maps (we omit the derivation
here, it is similar to the derivation of $nGLS$):
\begin{eqnarray*}
skewed-nGLS(a,p,x) &=&
\frac{(a-p)+\sqrt{(p-a)^2+4ax}}{2a}~~~~~~~~~~~~~~~~~~~~~~~~~~~
0\leq x < p\\\\
  &=& \frac{(1+a-p)-\sqrt{(p-a-1)^2+4a(1-x)} }{2a}~~~~~~~~~ p \leq x \leq 1
\end{eqnarray*}
where $0 \leq p \leq 1$ and for a given `$p$', we have $0 < a \leq
p$. Figure~\ref{fig:figsngls1} shows the plot of skewed-nGLS for a
few values of `$p$' and `$a$'. Note that as $a \rightarrow 0$ in
each case, the $skewed-nGLS$ tends to the skewed tent-map.
\par Skewed-nGLS seems to exhibit Robust Chaos as depicted in the
bifurcation diagram (Figure~\ref{fig:figsngls2}). We show the
bifurcation diagram for values outside the permissible range of
`$a$' in Appendix A.
\par The Lyapunov exponents of skewed-nGLS can be easily derived by observing
that the invariant density of the skewed-nGLS is uniform
distribution. We can then use the Ergodic theorem to derive the
Lyapunov exponents:
\begin{eqnarray*}
\lambda(a,p) &=& \frac{ (a+p-1)^2 log(a+p-1)^2 - (a-p+1)^2
log(a-p+1)^2}{8a}  \\
& & + ~\frac{ (-a+p)^2 log(-a+p)^2  - (a+p)^2 log(a+p)^2 }{8a}
+\frac{1}{2}
\end{eqnarray*}
It can be seen that the Lyapunov exponents are always positive for
$0 < a \leq k$ where $k = min\{p, 1-p \}$.
\newpage
\begin{figure}[!hbp]
\centering
\includegraphics[scale=.6]{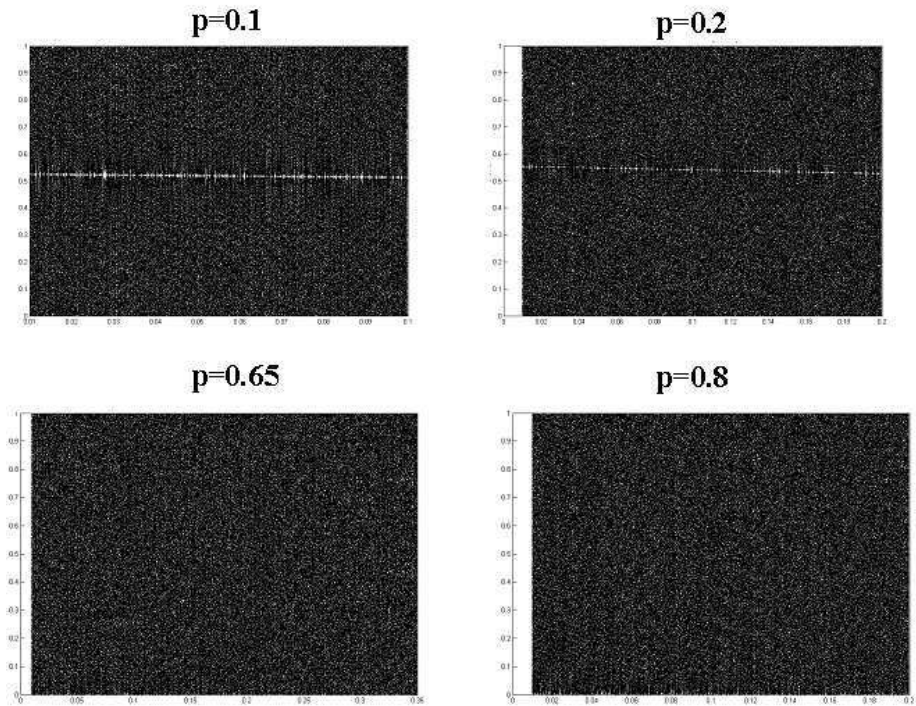}
\caption{Skewed-nGLS exhibits Robust Chaos as shown by the
bifurcation diagram for a few `$p$' values. The white streak for
$p=0.1$ and $p=0.2$ disappear when iterated for a long enough
time.} \label{fig:figsngls2}
\end{figure}
\section{Joint Entropy Coding and Encryption using Skewed nGLS}
We are now ready to propose a new algorithm for joint entropy
coding and encryption using skewed-nGLS. We propose using the
parameter `$a$' as a private key. Encoding and decoding will be
now done with $nGLS(a,p,x)$. The algorithm is exactly the same as
before (Section 5.1). Here `$p$' is the source probability of the
occurrence of the alphabet $0$ (or $A$) and $1-p$ corresponds to
the probability of occurrence of $1$ (or $B$). The key space for
`$a$' is $(0, k]$, where $k = min\{p, 1-p \}$.

\subsection{Shannon's desired sensitivity of the key parameter
`$a$'} We shall demonstrate that our method achieves Shannon's
desired sensitivity of the key parameter `$a$'. To this end, let
us assume a precision of $\delta = 10^{-12}$. Consider two keys
$a_1$ and $a_2 = a_1 + \delta$. We chose a random initial
condition $IC = 0.79193703742704$ and forward iterate
$nGLS(a,p=0.5,x)$ for the two keys $a_1$ and $a_2$ with the same
$IC$. We compare the symbolic sequences of the two trajectories.
\begin{figure}[!hbp]
\centering
\includegraphics[scale=.6]{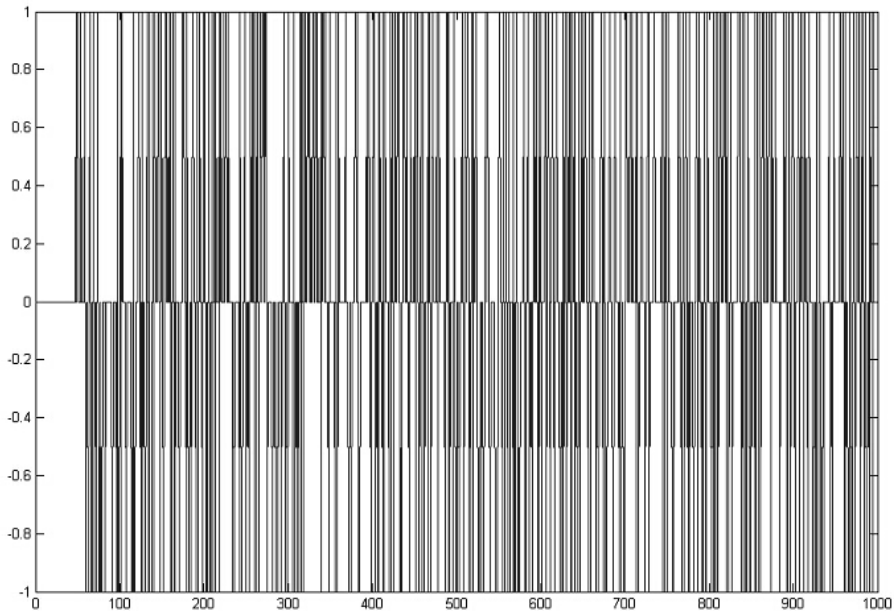}
\caption{Shannon's desired sensitivity of the key parameter `$a$'.
The difference in the symbolic sequences of $nGLS(a_1,p=0.5,x)$
and $nGLS(a_2,p=0.5,x)$ is plotted across iterations. As it can be
seen, after around 45-47 iterations, they are uncorrelated. $a_2 =
a_1 + \delta$, $\delta = 10^{-12}$.} \label{fig:figjac1}
\end{figure}
\par As it can be seen from Figure~\ref{fig:figjac1}, the two
symbolic sequences are uncorrelated after $45-47$ iterations. For
a $\delta = 10^{-7}$, they become uncorrelated after 35
iterations. This shows good sensitivity. We have also found that
the same is true for various values of `$p$' (not just limited to
$p=0.5$). The method we propose is to append $random$ data for the
first 50 bits, since it is possible for an eavesdropper to decode
correctly up to the first 50 bits with an arbitrary guessed key.
We could have a 40-bit key, with a maximum of $k \times 10^{12}$
keys which `$a$' can take. Here the maximum value of `$k$' is
$0.5$. The value of the `$k$' depends on `$p$'. Remember, $k =
min\{p, 1-p \}$ and $0 < a \leq k$.
\subsection{Advantages of our method}
\begin{enumerate}
\item One time private key (`$a$') transmission. %
\item Large key space -- a maximum of half a billion keys. %
\item Sensitivity to key parameter due to "Robust Chaos". %
\item No windows, no attractive periodic orbits. Well suited for
hardware/analog circuit implementation. \item Overhead is very
small $< 100$ bits (50 bits  of random data +
40 bits of key). %
\item Is skewed-nGLS ergodic? We believe it is, though we have no
proof at this point. Ergodicity is an important property which
would imply mixing or independence, desirable for encryption.
\end{enumerate}
But, we need to find out what happens to the compression ratio?
\subsection{Compression ratio efficiency}
The compression ratio is {\it not} optimal, as it depends on
`$a$'. For values of `$a$' close to 0 yields us closer to the
standard AC and is optimal, but it offers bad encryption. However,
we note that the total compression ratio is preserved. The {\it
length} of the compressed data is scrambled (unlike traditional AC
where the length is the same). This implies that two sequences
with the same probability may not get the same length. We could
swap modes to create efficient scrambling of lengths. The swapping
sequence could be based on a key, but unlike Grangetto's method,
we can make the swapping sequence a public key and retain `$a$' as
the private key.
\section{Summary and Future Research Directions}
We have established that GLS coding is equivalent to Arithmetic
Coding and hence Shannon optimal. GLS is a chaotic, ergodic and
measure-preserving discrete dynamical system. We have provided a
motivation for the use of GLS in a framework of joint entropy
coding and encryption. We have derived measure preserving
piece-wise non-linear GLS (nGLS) and their skewed cousins
(skewed-nGLS). nGLS and skewed-nGLS exhibit positive Lyapunov
exponents and ``Robust Chaos'', both of which are necessary for
{\it strong} encryption. We have proposed a method for joint
compression and encryption using skewed-nGLS which exhibits a
reasonably large key space and one-time private key transmission.
Shannon's desired sensitivity of keys (due to Robust Chaos) has
been demonstrated for this method. We note that the total
compression ratio is preserved and the {\it length} of the final
interval (in which the tag is going to lie) is scrambled.

\par However, we {\it don't} claim that our method is secure or optimal. We need to perform rigorous
cryptanalysis (known plain-text attack, known cipher-text
plain-text pair attack, differential cryptanalysis etc.) We also
wish to perform analysis on compression ratio distribution of
various messages and quantify the loss of optimality of
compression ratio and how it can be minimized. Randomizing the
length information in an efficient manner while still retaining
compression efficiency is important and here swapping modes based
on publicly announced sequences would play a role. Issues related
to computational precision and decoder complexity need to be
addressed. Last but not the least, we need to perform rigorous
testing, especially on long sequences.

\par We hope that we have provided enough hope for utility of Chaos and Robust Chaos in
communications by indirectly showing that the popular Arithmetic
Coding algorithm is in fact a dynamical system exhibiting Chaos.
\section*{Acknowledgements}
We would like to express our sincere gratitude to the Department
of Science and Technology for funding the Ph.D. fellowship program
at the National Institute of Advanced Studies. We would like to
thank Indian Institute of Science (specifically the CSA and MATH
departments) for allowing us to take courses. We would also like
to thank Professor Vidhu Prasad of the University of Massachusetts
Lowell for introducing some of these ideas to us, and Sajini
Anand, Suhail Ullal and David McCandlish for their valuable
discussions.

\par A version of this work was recently presented at the National
Conference on Mathematical Foundations of Coding, Complexity,
Computation and Cryptography IISc., Bangalore, July $20-22$nd,
2006.

\newpage
\section*{Appendix}
\appendix
\section{Bifurcation diagrams of $skewed-nGLS(a,p,x)$ for various values of `$p$' for large ranges of `$a$'}
We shall plot here the various bifurcation diagrams of
$skewed-nGLS(a,p,x)$ for various values of `$p$' and for values of
of the bifurcation parameter `$a$' outside the range $(0, k]$
where $k = min\{p, 1-p \}$.
\begin{figure}[!hbp]
\centering
\includegraphics[scale=.7]{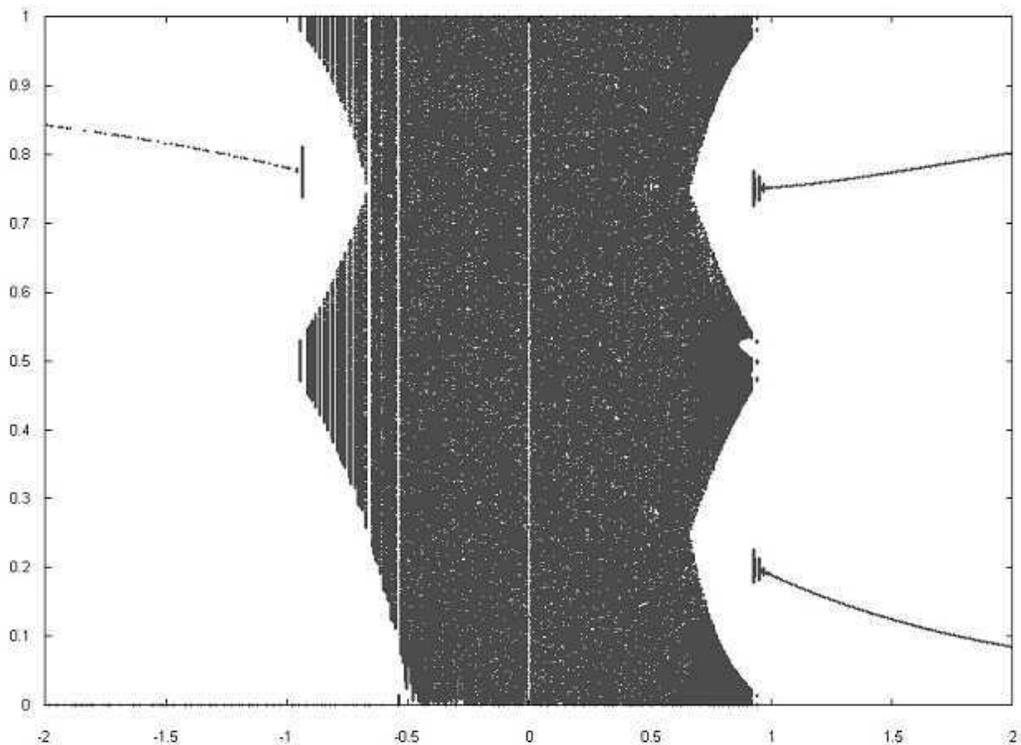}
\caption{Bifurcation diagram of $nGLS(a,p=0.5,x)$.}
\label{fig:figbif1}
\end{figure}
\begin{figure}[!hbp]
\centering
\includegraphics[scale=.8]{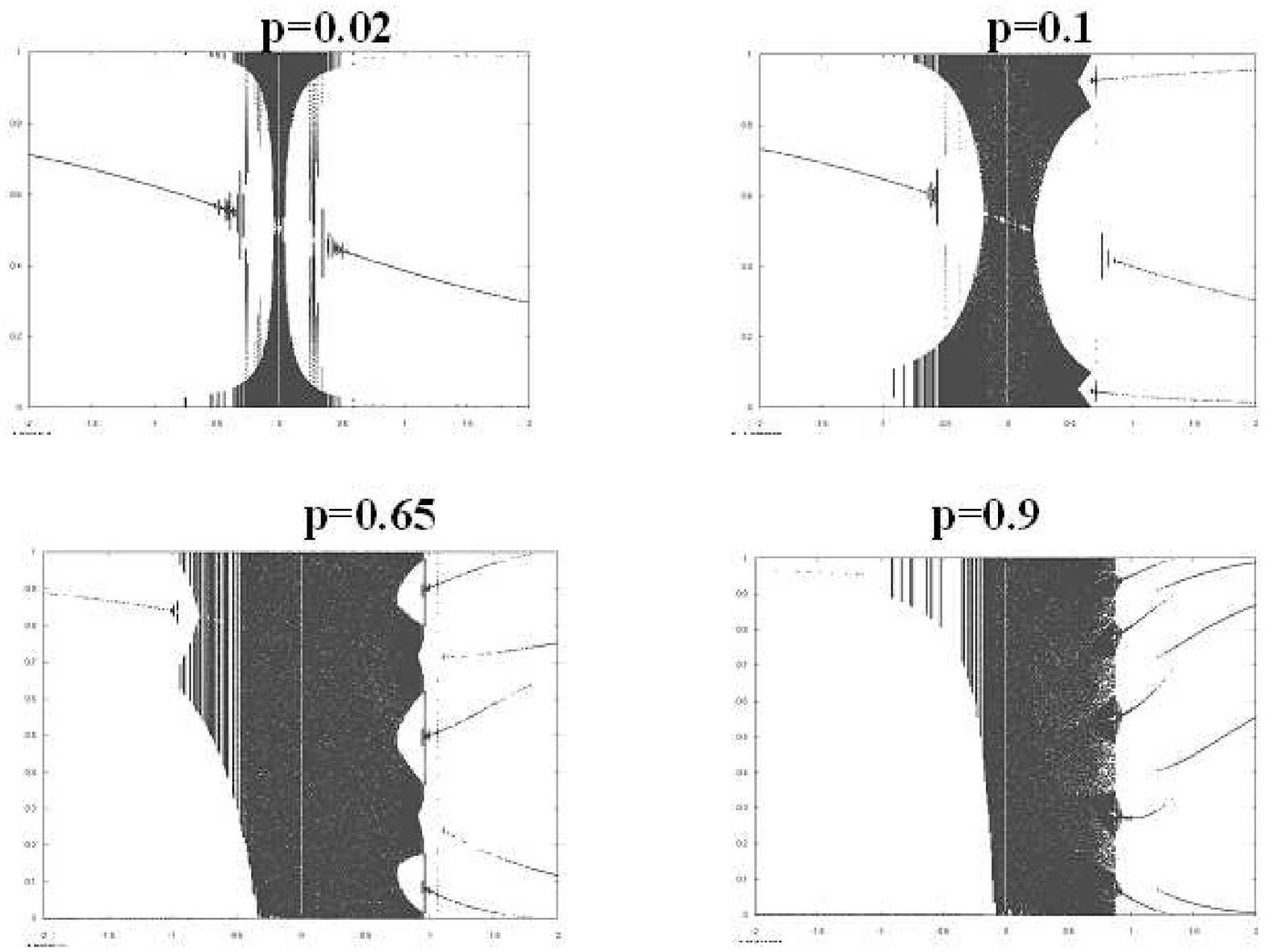}
\caption{Bifurcation diagram of $nGLS(a,p,x)$ for various values
of `$p$'. One can notice the white streak in the center of the
plot for small values of `$p$' (0.02 and 0.1). This white streak
disappears when computed for large number of iterations.}
\label{fig:figbif2}
\end{figure}

\begin{thebibliography}{1}
\bibitem {Shannon48} C E Shannon, A Mathematical Theory of
Communication, Bell System Technical Journal, vol. 27, pp.
379--423. (1948)

\bibitem{Salomon}
D Salomon, Data Compression: The Complete Reference, New York: 2nd
edn. Springer-Verlag. (2000)

\bibitem{Huffman}
D A Huffman, A method for the construction of minimum-redundancy
codes, Proceedings of the I.R.E., pp. 1098--1102. (sep. 1952)

\bibitem{Aligood}
K T Alligood, J A Yorke, T D Sauer, Chaos: An Introduction to
Dynamical Systems, Springer-Verlag Inc. (1997)

\bibitem{Dajani}
K Dajani, C Kraaikamp, Ergodic Theory of Numbers, Carus
Mathematical Monographs, 29. Mathematical Association of America,
Washington, DC. (2002)

\bibitem{Rissanen}
J J Rissanen and G G Langdon, Arithmetic Coding, IBM Journal of
Research and Development, vol. 23, no. 2,  pp. 146-162. (Mar.
1979)

\bibitem{Sayood}
K Sayood, Introduction to Data Compression, Morgan Kaufmann.
(1996)

\bibitem{Luca}
M B Luca, A Serbanescu, S Azou, G Burel, A New Compression Method
using a Chaotic Symbolic Approach,
www.univ-brest.fr/lest/tst/publications/pdf/comm04\_compression\_chaos.pdf

\bibitem{Grangetto}
M Grangetto, A Grosso, E Magli, Selective Encryption of JPEG2000
Images by Means of Randomized Arithmetic Coding,  IEEE 6th
Workshop on Multimedia Signal Processing, Sienam, Italy, pp.
347-350. (Sep. 2004)

\bibitem{Wen}
J G Wen, H Kim, J D Vilasenor, Binary Arithmetic Coding Using
Key-Based Interval Splitting, IEEE Signal Processing Letters, vol.
13, no. 2. (Feb. 2006)

\bibitem{Shannon49}
C E Shannon, Communication Theory of Secrecy Systems, Bell System
Technical Journal, vol. 28, pp. 656-715. (1949)

\bibitem{Kocarev}
I Kocarev, Chaos Based Cryptography, A Brief Overview, IEEE
Circuits and Systems Magazine, vol. 1, no. 3, pp. 6-21. (2001)

\bibitem{Banerjee}
S Banerjee, J A Yorke, C Grebogi, Robust Chaos, Physics Review
Letters, vol. 80, no. 14. (Apr. 1998)

\bibitem{Barreto}
E Barreto, B R Hunt, C Grebogi, J A Yorke, From High Dimensional
Chaos to Stable Periodic Orbits: The Structure of Parameter Space,
Physics Review Letters, vol. 78, no. 24. (Jun. 1997)

\bibitem{Andrecut}
M Andrecut, M K Ali, Robust Chaos in Smooth Unimodal Maps,
Physical Review E, vol. 64, no. 025203. (Jul. 2001)

\bibitem{Andrecut2}
M Andrecut, M K Ali, Example of robust chaos in a smooth map,
Europhyics Letters, vol. 54, no. 3. (May 2001)

\bibitem{Mahesh}
M C Shastry, N Nagaraj, P G Vaidya, The B-Exponential Map: A
Generalization of the Logistic Map, and Its Applications In
Generating Pseudo-random Numbers, arXiv.org:cs/0607069. (Jul.
2006)
%
%\bibitem{Walters}
%P Walters,  An Introduction to Ergodic Theory, New York:
%Springer-Verlag. (2000)
%
%
\end{thebibliography}
\end{document}